# Dynamic Directional Routing of Freight in the Physical Internet

Sahrish Jaleel Shaikh[1], Praveen Muthukrishnan[1], Yijun Lai[1], Benoit Montreuil[1]

1. H. Milton Stewart School of Industrial & Systems Engineering,
Georgia Institute of Technology, Atlanta, United States
Corresponding author: sahrish.shaikh@gatech.edu

**Abstract**

The Physical Internet (PI) envisions an interconnected, modular, and dynamically managed logistics system inspired by the Digital Internet. It enables open-access networks where shipments traverse a hyperconnected system of hubs, adjusting routes based on real-time conditions. A key challenge in scalable and adaptive freight movement is routing—determining how shipments navigate the network to balance service levels, consolidation, and adaptability. This paper introduces directional routing, a dynamic approach that flexibly adjusts shipment paths, optimizing efficiency and consolidation using real-time logistics data. Unlike shortest-path routing, which follows fixed routes, directional routing dynamically selects feasible next-hop hubs based on network conditions, consolidation opportunities, and service level constraints. It consists of two phases: (1) area discovery, which identifies candidate hubs, and (2) node selection, which determines the next hub based on real-time parameters. This paper advances the area discovery phase by introducing a Reduced Search Space Breadth-First Search (RSS-BFS) method to systematically identify feasible routing areas while balancing service levels and consolidation. The proposed approach enhances network fluidity, scalability, and adaptability in PI-based logistics, advancing autonomous and sustainable freight movement.

## 1. Introduction

The Physical Internet (PI) reimagines logistics as a seamlessly connected, modular, and dynamically managed system, where shipments flow through an interconnected multi-party network of hubs, much like data packets in the Digital Internet [1]. Current logistic routing methods, such as shortest-path routing or fixed-route planning, struggle to accommodate the dynamic nature of real-world supply chains, where demand fluctuations, congestion, and operational constraints necessitate more flexible decision-making.

In the Digital Internet, data packets do not follow a fixed, pre-determined route. Instead, they traverse a dynamic set of intermediate nodes, leveraging real-time network conditions to optimize performance. Dynamic directional routing, inspired by opportunistic routing in wireless networks [2], is a transformative approach to logistic operations. In opportunistic routing, each packet dynamically selects the next relay node based on real-time network conditions, optimizing the overall path and performance. Similarly, in logistics, dynamic directional routing involves real-time choice of intermediate hubs or relay points for shipments of modular containers from tote and pallet sizes to cargo sizes, guided by current network states such as hub congestion, transportation capacity, and service level requirements. This paradigm shift moves away from fixed routes, embracing a flexible, data-driven methodology that continuously adjusts as shipments progress through the network.

This paper introduces a novel framework for dynamic directional routing [3] within the context of the Physical Internet (PI), leveraging principles of modularity, hyperconnectivity, and openness. By integrating real-time data streams on traffic conditions, hub activity levels, and container availability, our proposed routing protocol enables adaptive decision-making at every stage of a shipment's journey. Specifically, we develop and evaluate an innovative protocol that organizes logistic networks into sector-based structures. This protocol dynamically selects optimized paths, ensuring efficient shipment consolidation and minimal delays while keeping high service levels.

The contributions of this paper are threefold. First, we conceptualize and formalize the adaptation of opportunistic routing principles to logistic networks, highlighting their potential to enhance efficiency and resilience. Second, we design and implement simulation-based evaluations of the proposed protocol, benchmarking its performance against traditional static routing methods. Third, we explore the broader implications of dynamic directional routing for achieving sustainable and resilient supply chains. The remainder of this paper is organized as follows. In Section 2, we review the relevant literature and highlight how our approach connects to established streams in logistics and telecommunications. Section 3 presents the conceptual framework for dynamic directional routing in the Physical

Internet, including the detailed Reduced Search Space BFS (RSS-BFS) algorithm. Section 4 describes the simulation setup and performance evaluation results, contrasting our proposed directional routing with the baseline shortest-path approach. Finally, Section 5 concludes the paper, discussing insights, limitations, and potential future research directions for advancing the proposed framework.

## 2. Related Works

Freight routing in logistics has been extensively studied [4,5], with traditional approaches focusing on static and deterministic path optimization. Methods such as the Traveling Salesman Problem (TSP) [6] and Vehicle Routing Problem (VRP) [7,8] have provided foundational tools for route planning, often emphasizing cost minimization and efficiency. However, these methods typically assume fixed conditions and are ill-suited for dynamic, real-world scenarios where disruptions, demand fluctuations, and network changes are common [9,10].

Dynamic routing, by contrast, introduces flexibility by leveraging real-time data to adjust routes. Early studies in dynamic vehicle routing explored adaptive strategies based on traffic conditions and delivery priorities [11,12]. These approaches demonstrated significant improvements in responsiveness and efficiency but often lacked scalability and integration with broader logistics networks [13]. More recent advancements, such as real-time fleet management systems, have integrated GPS tracking and IoT-enabled sensors to further enhance dynamic routing capabilities [14]. Parallel developments in telecommunications, particularly in opportunistic routing for wireless networks [15], offer valuable insights for logistics. Opportunistic routing techniques, such as ExOR [16] and MORE [17], prioritize dynamic relay selection based on instantaneous network conditions, thereby improving throughput and reliability [18]. These methods highlight the potential of adaptive, context-aware decision-making to optimize performance in complex, decentralized systems. Drawing inspiration from these principles, this paper applies similar concepts to logistics, focusing on dynamic hub selection and sector-based routing to address the unique challenges of physical goods movement.

While prior research has explored elements of dynamic routing in both logistics and telecommunications [15][19], the application of opportunistic routing principles to freight movement remains limited. This paper builds on insights from opportunistic routing in telecommunications and adapts them to the Physical Internet framework [1], developing a novel protocol that leverages real-time data for dynamic shipment routing. In doing so, it advances the development of sustainable, resilient, and efficient supply chain networks [20].

## 3. Conceptual Framework and Methodology

As established in Section 1, Directional Routing in the Physical Internet (PI) enables adaptive decision-making, allowing shipments to move through a dynamic network of hubs rather than following a pre-determined shortest path. The first phase of this routing process, Area Discovery, determines a set of candidate hubs that a shipment may pass through, much like the forwarding set in the Digital Internet.

Traditional shortest-path routing predetermines the entire path upfront, making it inflexible and often leading to suboptimal consolidation. Directional Routing, however, selects only the next hub dynamically, requiring a method to discover candidate hubs efficiently at each decision point. Area discovery ensures that a shipment has multiple viable routing options, balancing consolidation, service constraints, and computational efficiency. The challenge in Area Discovery is limiting unnecessary path exploration while ensuring that enough viable hubs are considered. To address this, we develop a Reduced Search Space Breadth-First Search (RSS-BFS) Algorithm, which discovers candidate hubs while respecting service constraints and consolidation opportunities.

### 3.1 Problem Definition: Candidate Hubs Selection

Routing freight across logistic networks often demands more than a simple shortest-path solution, particularly when dealing with dynamically and stochastically evolving inter-hub travel times and intra-hub dwell times, multiple service-level agreements and the need for shipment consolidation. We model the network as a graph G(H,E), where the set H comprises all hubs, and each edge in E captures feasible connections between these hubs. Within this network, a shipment must travel from an origin hub $H_s$ to a destination $H_d$, subject to a maximum allowable travel time or cost constraint determined by the shipment's service level.

The objective is to determine a subset of hubs, denoted as $H_c$, that qualify as candidate next hops by meeting specific feasibility criteria while avoiding unnecessary search space expansion. These candidate hubs form a filtered set of routing options that align with the shipment's direction of travel and service-level requirements, while also enhancing consolidation and cost efficiency. Since directionality, time constraints, and network connectivity influence which hubs qualify, the selection process must ensure that only operationally relevant hubs remain in $H_c$, balancing efficiency with the practical constraints of supply chain operations.

In identifying candidate hubs, several key criteria must be met. First, candidate hubs should align with the general direction of the shortest path, ensuring that deviations do not cause excessive increases in transportation or handling time. Second, hubs must allow feasible routing within the shipment's maximum travel-time constraint to ensure on-time delivery. Third, hubs should support consolidation opportunities to minimize unnecessary shipment splitting. Finally, only hubs that are physically and administratively capable of handling transshipments should be considered. These criteria necessitate a targeted algorithmic approach to efficiently filter out non-viable hubs early, reducing unnecessary exploration of impractical or inefficient paths.

### 3.2 Reduced Search Space BFS (RSS-BFS) Algorithm

The RSS-BFS Algorithm is designed to minimize computational overhead while still discovering all relevant hubs that could potentially improve the routing plan. Traditional BFS explores neighbors of a node indiscriminately, which can rapidly inflate the number of possible paths, especially in dense networks. By contrast, RSS-BFS employs a suite of pruning rules aiming to ensure that only viable candidates continue to be explored. The core principles are:

- *Directional Filtering*
  Before we expand from a node, we calculate the bearing from the origin to a reference point (often involving the destination) in order to determine if a neighboring hub leads in the "positive" direction of travel. This filter significantly reduces exploration of irrelevant or backwards-moving routes.
- *Positive Sector Expansion*
  Instead of exploring the full 360° around a given node, we define a "sector" centered around the computed bearing (e.g., ±50°). Any hub whose bearing from the current node falls outside this angular range is discarded from further consideration, mitigating unnecessary branching.
- *Search Space Pruning*
  As the algorithm proceeds, it continuously checks cumulative travel times (and any additional handling times) against the service-level constraint. If incorporating a hub pushes the path beyond the allowable time, or if no viable continuation to the destination exists from that hub, the algorithm prunes that branch immediately.

By cutting out large swaths of the network that do not contribute to a feasible route, RSS-BFS significantly reduces the search space, thus lowering computational time and memory usage. The focus on directionality and time-based feasibility ensures that only pertinent routing options are considered, improving the overall quality of solution paths.

### 3.3 Algorithm Implementation

The **RSS-BFS Algorithm** can be broken down into four primary stages, each designed to refine the set of possible next-hop hubs in a structured manner while respecting the service-level envelope.

### Step 1: Compute Shortest Path and Initial Bearing

The process starts by determining a baseline shortest path from the origin $H_s$ to the destination $H_d$ using a classical algorithm such as Dijkstra's. From this path, the algorithm identifies the first hub and calculates a bearing that connects the origin with a midpoint representing the first hub and the destination. This bearing serves as an anchor for the directional filter to follow. The use of a shortest-path baseline ensures that the default direction of travel is grounded in a feasible route, though the algorithm remains flexible enough to deviate slightly if needed.

### Step 2: Filter Neighbors in the Positive Routing Sector

With the reference bearing established, the algorithm examines the immediate neighbors of the current node. For each neighboring hub, it calculates the bearing relative to the current hub's position. If that bearing deviates more than a predefined angular limit from the anchor bearing, the neighbor is excluded. This focuses the search on nodes that nudge the shipment closer to the destination, rather than dispersing the search radially in every possible direction.

### Step 3: Perform BFS Expansion with Service-Level Constraint

Traditional BFS is then conducted within the limited set of neighbors that survived the directional filter. However, each time the algorithm moves from one hub to another, it evaluates cumulative travel and handling times. If adding a hub pushes the path beyond the service-level constraint, that path is pruned immediately to prevent further expansion along an infeasible route. In some cases, if no valid candidates remain within the positive sector, the algorithm may recalculate bearings using alternative anchor points (for example, the bearing from $H_s$ to the first hub in the shortest path directly) to avoid prematurely eliminating potential paths. If no feasible path remains after recalibration, it indicates that the freight will be late with no alternative routing options. In such cases, the system automatically considers lead-time extensions to prevent deadlocks. Additionally, special cases where no feasible solution exists are logged, and responsible agents are notified to take corrective action.

**Step 4: Select the Final Set of Candidate Hubs**

When the BFS expansion concludes—or no further expansions can be made without violating the time constraint—the remaining unpruned hubs become the candidate set $H_c$. These hubs each offer a realistic next-hop option in the path from $H_s$ to $H_d$ while ensuring the route stays within maximum allowable travel times. Consequently, rather than dealing with the entire network, downstream decision-making processes only consider a carefully curated list of hubs, making both route optimization and scheduling far more tractable.

By combining directional filtering, service-level constraints, and a sector-based BFS expansion, the RSS-BFS algorithm drastically narrows the set of potential routes without sacrificing feasibility or violating operational requirements. The final output, a concise collection of candidate hubs, provides a robust foundation for subsequent routing and scheduling decisions that seek to optimize costs, maintain quality of service, and exploit consolidation opportunities. Figure 1 provides images to showcase a subset of the many possible alternate routes generated by the refined RSS-BFS algorithm, highlighting how different neighbor-selection strategies and parameters can create multiple valid paths toward the same destination.

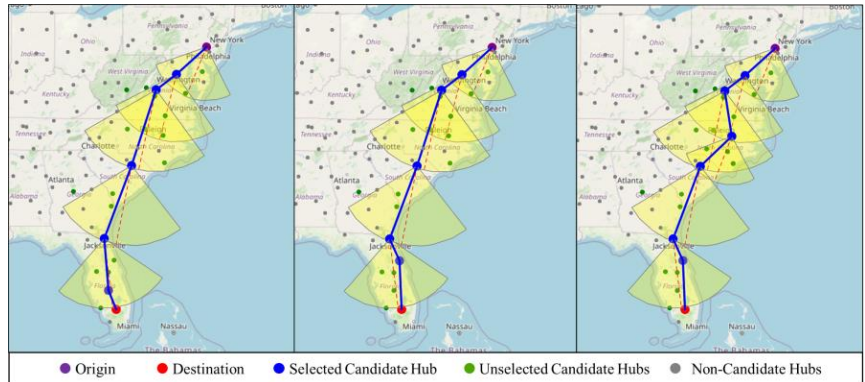

Figure 1: Alternative Routes Generated by RSS-BFS from a Hub near New York to a hub in South Florida

## 4. Performance Evaluation and Simulation Results

We evaluate the proposed *directional routing* approach against the *traditional shortest-path routing* (baseline). Both methods use the same global consolidation logic, ensuring that any observed differences can be attributed primarily to the routing strategy itself. We model and use a transportation network in the Southeastern United States, with multiple origin hubs and two primary destination hubs for shipment routing. Each shipment has an associated service level (1, 2, or 3), translating into a delivery deadline of 24, 48, or 72 hours, respectively. Shipments are gradually generated and introduced into the simulation at different times within a specified timeframe (e.g., over a 12-hour period). This approach ensures that shipments do not all enter the system simultaneously, allowing for a more realistic flow of freight through the network. Each shipment's origin, destination, service level, and creation time are randomly selected to reflect heterogeneous traffic.

The simulator models both vehicle movements and freight handling processes. It processes truck dispatch events, including capacity constraints and routing decisions, while also managing shipment arrivals, hub operations such as consolidation, and freight-specific handling times at intermediate hubs. Since multiple shipments can be transported within a single truck, the model explicitly distinguishes between truck-level operations and shipment-level events to ensure accurate representation of both vehicle movements and freight flow.

In the baseline (Traditional Shortest Path), each shipment follows the precomputed shortest travel-time path, moving from hub to hub until reaching its destination. In the directional routing case, upon a shipment's arrival at a hub, the next-hop candidates are dynamically determined by focusing on hubs that lie within ±50° of the bearing toward the destination. Among these candidates, the next hop is chosen based on travel time and global consolidation considerations (i.e., grouping shipments headed in the same general direction).

In both modes, the same consolidation protocol applies: the simulator checks whether enough shipments are ready for dispatch based on waiting time thresholds or capacity triggers. Trucks are called on-demand and arrive within a predefined time window (e.g., X hours). The time required for loading depends on the number of shipments being loaded. Once a truck is dispatched, the shipments it carries are removed from the hub's queue until they arrive at the next node. As shown in Table 1, we observe the following trends in the simulation results:

- **Number of Trucks Dispatched**: Directional routing consistently reduces the number of trucks required for shipment delivery, with reductions of up to 18% compared to the baseline. This is particularly evident in higher-demand scenarios, where fewer trucks translate to potential savings in labor and fleet operating costs. This efficiency is illustrated in Figure 2, which highlights the consolidation benefits of directional routing compared to the baseline approach.

- **Delivered vs. Undelivered Shipments**: Both baseline and directional routing deliver the majority of shipments within their deadlines; however, routing decisions can differ significantly in intermediate steps.
- **Total Miles Traveled:** Under lower demand scenarios, both baseline and directional routing exhibit similar performance. Under certain scenarios (e.g., 600 shipments, low demand), directional routing travels marginally more miles (+0.3%). However, in moderate-to-high demand cases, it can slightly reduce total distance (up to -4.6%), reflecting more efficient corridor consolidation, as shown in Figure 2. Directional routing generally reduces the number of trucks by up to 18%, especially at higher demand. Fewer trucks can translate to lower labor and fleet operating costs, although this benefit must be weighed against the slight additional cost in total mileage observed in some cases.
- **Delivery Times and Path Diversity**: The shortest-path approach consistently relies on the minimum travel time sequence of hubs, while the directional approach may explore alternate but still feasible routes, especially useful under tight deadlines or capacity constraints and is further illustrated in Figure 3, where different routing paths are visualized for various origin-destination pairs.

Table 1: Comparing KPIs of Baseline and Directional Routing for Low, Moderate, and High Demand Scenarios

| Demand Level | Number of Shipments | Routing Scheme | Shipments Delivered On-time | Trucks Dispatched (Δ%) | Total Miles (Δ%) |
|---|---|---|---|---|---|
| **Low** | 600 | Baseline | 100% | 485 | 91,868 |
|  |  | Directional | 100% | 433 **(-10.7%)** | 92,102 **(+0.3%)** |
| **Moderate** | 1000 | Baseline | 100% | 533 | 100,638 |
|  |  | Directional | 100% | 472 **(-11.5%)** | 100,429 **(-0.2%)** |
|  | 1200 | Baseline | 100% | 657 | 126,986 |
|  |  | Directional | 100% | 538 **(-18.1%)** | 122,545 **(-3.5%)** |
| **High Demand** | 2000 | Baseline | 100% | 721 | 139,301 |
|  |  | Directional | 100% | 589 **(-18.3%)** | 132,902 **(-4.6%)** |

Overall, these results confirm that directional routing can perform on par with or better than the baseline in situations where a small angular deviation helps group shipments more effectively, reduce partial truck loads, and exploit multiple viable paths to the final destination. Further investigations might assess how varying the angular width or consolidation threshold affects delivery times and resource usage.

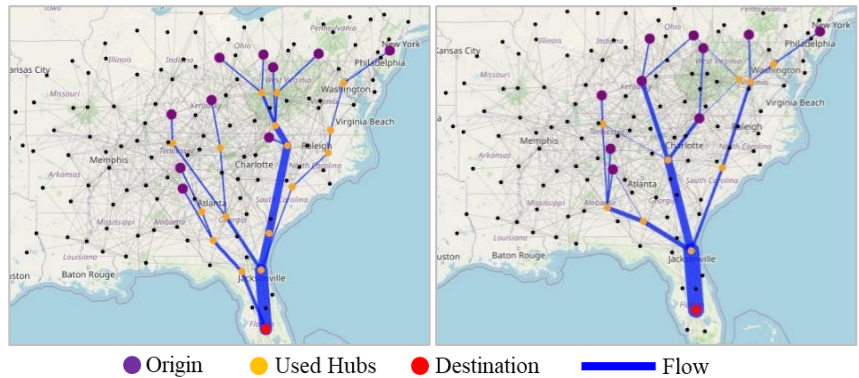

Fig 2: Path Selection and Consolidation Efficiency: Baseline (Left) and Directional Routing (Right)

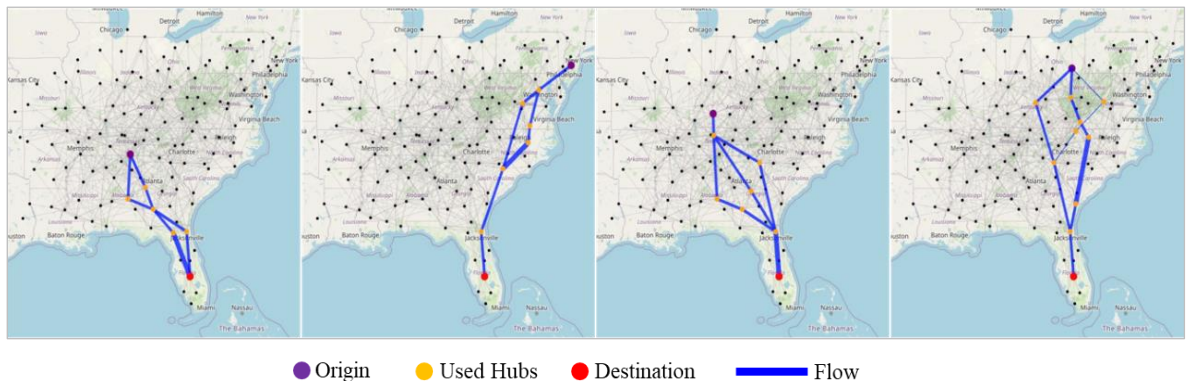

Fig 3: Alternate Routes in Directional Routing – Visualizing Different Origin-Destination Pairs

# 5. Conclusion

In this paper, we presented a novel directional routing framework tailored for Physical Internet (PI) logistics networks, adapting core ideas from opportunistic routing in wireless communications. Through comparative simulation experiments against a traditional shortest-path baseline, the results indicate that directional routing consistently meets deadlines across varying demand levels, while potentially reducing truck usage and improving corridor-based load factors—particularly in moderate-to-high demand scenarios. While the proposed directional routing framework demonstrates clear potential, several avenues remain open for further exploration and enhancement:

- *Resilience and Disruption Handling:* Investigating hub or edge failures will help validate how quickly the algorithm can reroute shipments without compromising service levels.
- *Advanced Dispatch Protocols:* Integrating priority-based or schedule-based dispatch triggers could further optimize resource usage and adapt to fluctuating demand patterns.
- *Scalability*: Examining performance on larger or more densely connected networks would confirm the solution's viability for real-world mega-regional or global logistics contexts.
- *Real-Time Data*: Incorporating dynamic information such as traffic conditions or temporary hub constraints would allow continuous recalibration of angular widths and consolidation thresholds.
- *Multi-Criteria Optimization*: Balancing various objectives, such as minimizing cost and reducing emissions would align directional routing more closely with sustainability and operational goals in modern supply chains.

Overall, directional routing positions logistics operations closer to the Physical Internet vision, providing a flexible, data-driven mechanism for determining routes and consolidating shipments in real-time. By preserving service-level requirements while leveraging dynamic corridor-based searches, this framework enables freight networks to become more agile, scalable, and resource-efficient. Further explorations in resilience, adaptive angles, and multi-criteria strategies will help fully realize the potential of this approach.